# ELECTRIC AND MAGNETIC DIPOLE MOMENTS OF FREE NANOCLUSTERS

Walt A. de Heer[1] and Vitaly V. Kresin[2]

[1]School of Physics, Georgia Institute of Technology, Atlanta, GA 30332–0430, USA.
E-mail <deheer@electra.physics.gatech.edu>

[2]Department of Physics and Astronomy, University of Southern California, Los Angeles, CA 90089-0484, USA.
E-mail <kresin@usc.edu>

(*29 January 2009*)

## 1. Introduction

This article discusses the behavior and detection of electric and magnetic dipole moments in atomic and molecular nanoclusters. We concentrate on studies of free particles (i.e., clusters produced, probed, mass selected, and detected in beams (Pauly 2000)), because in such work the focus is on exploring the inherent physical properties of these nanoparticles. In this way, clusters can be studied with their size and composition precisely known, and with their features unperturbed by substrate interactions.

A fundamental characteristic of a molecule or nanoparticle is the arrangement and distribution of electrical charges within it, as well as that of electronic orbital and spin angular momenta. Electric and magnetic dipole moments are observables which directly reflect these distributions, and are therefore very useful experimental parameters for direct assessment of models and theories of particle structure, bonding, and internal dynamics. In addition, of course, polar and magnetic nanoclusters carry significant practical promise as building blocks for materials with novel magnetic and ferroelectric properties.

Although the principles underlying electric and magnetic ordering in nanoclusters may be distinct, the experimental signatures of such ordering in cluster beam work have a lot in common, in particular the patterns of beam deflection under the influence of externally applied fields. It is for this reason that they are discussed jointly in this article.

## 2. Definitions

As a starting point, it is appropriate to cite the textbook expression for the total electric dipole moment:

$$\vec{p} = \int \rho(\vec{r})\vec{r}\,d^3r \,. \tag{1}$$

Here $\rho(\vec{r})$ is the charge density (including both electrons and nuclei) at position $\vec{r}$ within the particle, and the integral is over the entire particle volume. It is important to keep in mind that the magnitude of the electric dipole moment is independent of the choice of the origin if, and only if, the particle on the whole is electrically neutral.

Electric dipole moments are usually quoted in Debye (D) units (1 D = $3.34 \times 10^{-30}$ C·m = 0.21 $e$·Å = 0.39 a.u.).

The corresponding definition of the magnetic dipole moment is

$$\vec{\mu} = \frac{1}{2c}\int \vec{r} \times \vec{J}(\vec{r})\,d^3r \,. \tag{2}$$

This is written in the Gaussian (*cgs*) system of units. $\vec{J}$ is the volume current density, and $c$ is the speed of light. Since current flow is associated with the motion, or velocity, of charges, the vector product is obviously related to the charges' angular momentum. The resulting quantum-mechanical expression for the magnetic moment of a system of electrons is

$$\hat{\vec{\mu}} = \frac{e\hbar}{2m_e c}\left(\hat{\vec{L}} + 2\hat{\vec{S}}\right). \tag{3}$$

The hats emphasize the fact that we're dealing with quantum-mechanical operators of the orbital and spin angular momentum. A similar equation would also describe nuclei; however, because the mass entering the denominator would be so much greater, the discussions and measurements of cluster magnetic dipole moments have so far focused on the dominant electronic contribution.

Magnetic dipole moments are usually quoted in Bohr magneton ($\mu_B$) units (1 $\mu_B$ = $e\hbar/2m_e c$ = $9.27 \times 10^{-21}$ erg/G = $9.27 \times 10^{-24}$ J/T).

As regards notation. one should be aware that the electric dipole moment is frequently denoted by $\vec{\mu}$, especially in the literature on molecular physics and physical chemistry, and also by $\vec{d}$; magnetic moments are sometimes denoted by $\vec{m}$.

## 3. Forces and Deflections Produced by External Fields

Many recent experimental measurements of the dipole moments of free clusters have been performed by the technique of beam deflection, sketched in Fig. 1. The main idea is that a narrow, accurately collimated beam of nanoclusters is directed through the region of a static inhomogeneous electric or magnetic field.

The potential energy of a permanent dipole in an external electric field is

$$U = -\vec{p} \cdot \vec{E}, \tag{4}$$

and by differentiating this one finds (Griffiths 1999) the force experienced by a polar particle:

$$\vec{F} = (\vec{p} \cdot \vec{\nabla})\vec{E}. \tag{5}$$

(It is assumed, of course, that the dimensions of the dipole are much smaller than the scale of spatial variation of the field.) The above equation reflects the fact that a homogeneous field will not exert a net force on the dipole (the forces on the negative and positive regions will balance), and therefore the apparatus field must be designed with a well-defined gradient. A similar equation can be written down for the magnetic force:

$$\vec{F} = (\vec{\mu} \cdot \vec{\nabla})\vec{B}. \tag{6}$$

Thus as a cluster particle passes between the electrostatic plates or magnet poles, two things happen. The field applies a torque to the dipole ($\tau = \vec{p} \times \vec{E}$ or $\vec{\mu} \times \vec{B}$), attempting to align it along the field lines; at the same time the particle is given an impulse by the collinear component of the field gradient.

Different quantum states (rotational, vibrational, electronic) of a cluster or a molecule usually possess different effective dipole moments, therefore one also needs to express the force in terms of the field-induced shift of state energy, i.e., in terms of the Stark (electric) or Zeeman (magnetic) shift. Thus for a state $|i\rangle$ with energy $\varepsilon_i(E)$ one has

$$\vec{F}_i = -\nabla \varepsilon_i(E) = \left(-\frac{\partial \varepsilon_i}{\partial E}\right)\left(\frac{\partial \vec{E}}{\partial z}\right) \tag{7}$$

(and analogously for magnetic fields and moments), where for conciseness we assumed that the field varies only along one coordinate, which is indeed the case for the most common experimental arrangements. In view of Eq. (5) the first factor in Eq. (7) can be viewed as the projection of the dipole moment in state $|i\rangle$ on the field direction, $p_{i,z}$ (or $\mu_{i,z}$). States with negative

Stark or Zeeman shifts (negative derivatives) are called high-field-seeking states, and those with positive shifts are called low-field-seeking.

One established field geometry comes from an equidistant arrangement of a set of electric or magnetic poles around a circle, in a cylindrical-type assembly (e.g., a "hexapole"). This creates a radial field near the cylinder axis (Parker and Bernstein 1989, Pauly 2000) which acts to defocus or refocus an axially collimated beam of molecules, depending on their quantum state, serving as a kind of thick lens for polar species.

For cluster beams, a much more common arrangement is a set of two parallel curved plates or poles producing the so-called Rabi "two-wire" deflecting field visible in Fig. 1, for which the field and its gradient are collinear and their product is approximately constant over a relatively large region around the beam axis (Ramsey 1956, Pauly 2000, Tikhonov et al. 2002). This type of field produces a sideways force, $F_z = p_z (\partial E/\partial z)$ or $F_z = \mu_z (\partial B/\partial z)$, and a corresponding sideways deviation of the particle:

$$d_z = a \langle p_z \rangle \frac{(\partial E/\partial z)}{m_c v^2}, \qquad (8)$$

where $\langle p_z \rangle$ is the time-averaged projection of the cluster dipole moment on the direction of the field, $a$ is a geometrical constant of the apparatus (encompassing such things as the length of the deflection plates and the flight distance from the deflection region to the detector; it is usually determined by calibrating the apparatus with a beam of well-characterized polarizable species), $m_c$ is the mass of the nanocluster, and $v$ is the beam velocity (one power of $v$ for the time spent in the field region, and another for the flight time to the detector; it is evident that accurate measurements require a good knowledge of beam velocity). An identical form can be written down for the magnetic case.

In a real experimental situation, one needs to average the above expression over all populated internal states and initial orientations of the entering particles, and convolute with the finite width of the incoming beam. Then finally one can extract information about the inherent electric or magnetic dipole moment of the nanocluster.

# 4. Beam Deflection and Broadening.
# "Rigid" and "Floppy" Polar Particles

*4.1 Beam Deflection*

It is evident from the foregoing that a critical bridge between the experimental signature and the intrinsic cluster moments *p* and *µ* lies in elucidating the average orientation of the former during passage between the electric plates or the magnetic poles. The two most commonly encountered parameters are the average deflection and the broadening of the beam.

The former corresponds to the average of $\langle p_z \rangle$ or $\langle \mu_z \rangle$ for the whole population of particular size clusters in the beam, or equivalently to the average orientation cosine, i.e., the expectation value of the angle between the molecular dipole axis and the field direction: $\langle \cos\theta \rangle = \langle p_z \rangle / p$. The averaging takes place over the flight time through the field and over the internal state distribution of the nanoclusters.

If the particles were in thermal equilibrium at some temperature *T*, the answer would be well known. For an ensemble of polar molecules in a gas in the presence of an external field, it was shown by Langevin (1905) that $\langle p_z \rangle = p[\coth x - 1/x] \equiv p\mathcal{L}(x)$, where $x \equiv pE/k_B T$ (or replace *p, E* by *µ, B*). $\mathcal{L}(x)$ is called the Langevin function. If $x \ll 1$, the polarization can be approximated by $\langle p_z \rangle = p^2 E / 3k_B T$, generally referred to as the Langevin-Debye law. The textbook derivation is usually done for a point dipole, but in *tour de force* calculations Van Vleck (1932) and Niessen (1929) proved that the result is valid (under realistic conditions) both in classical and in quantum mechanics for a general system possessing a dipole moment, with the understanding that $p^2$ represents a statistical mean over the phase space of the system.

Unfortunately, the situation for a transiting beam of polar nanoclusters is qualitatively different. Here the particles' state distributions are acquired in an earlier zero-field region (e.g., the rotational and vibrational temperature $T_{rot}$ produced at the beam source), and do change, i.e., re-thermalize, upon the relatively slow, adiabatic entry into the field. (Non-adiabatic, a.k.a. "Majorana," transitions between close-lying quantum states coupled by the field may complicate the situation further but will be neglected here.) Although the fact that low-field response should still scale with the ratio *pE/T* remains expected on dimensional grounds (Schnell et al. 2003), the numerical correlation between the magnitude of the dipole and the average force on the particle ensemble (and hence its deflection) is no longer universal.

As a result, experimental work and analysis have so far focused on, or sometimes simply assumed, certain treatable cases: (1) Clusters whose dipole moment rigidly rotates with the cluster's reference frame and does not interact with vibrations ("rigid" or "locked" dipole, or "strong magnetic anisotropy"); and (2) Clusters within which the orientation of the dipole can relatively easily readjust with respect to the particle axis ("floppy," "fluxional," or "superparamagnetic" cases). Situations with intermediate coupling strengths and anisotropy energies remain a challenge.

As mentioned above, in both cases 1 and 2 the contribution of the permanent dipole will scale as *pE/T* or *μB/T*, but a qualitative difference immediately arises: for "rigid" clusters the relevant *T* will be the rotational temperature of the beam, and for "floppy" clusters it will be their vibrational ("internal") energy. This is critically important, because for clusters in molecular beams the two temperatures do not at all have to be (and commonly aren't) equal.

For floppy clusters one normally assumes that their internal degrees of freedom are able to serve as a canonical heat bath for the dipole moment. For such systems one typically observes that the beam profile is deflected uniformly, as would a beam of particles possessing only an induced dipole moment: in Eq. (8) for $\langle p_z \rangle$ ($\langle \mu_z \rangle$) one has $\langle p_z \rangle = \chi E$ ($\langle \mu_z \rangle = \chi B$) with the linear susceptibility

$$\chi = \frac{p^2}{3k_B T_{\text{int}}}, \tag{9}$$

(or $\mu^2 / 3k_B T_{\text{int}}$) and the deflection is characteristically proportional to the square of the field intensity (since in deflection setups the field gradient and the field strength are proportional). For practical field strengths the linear Langevin-Debye limit of the susceptibility is adequate. Examples of such behavior will be presented in subsequent sections. It is important to appreciate that the thermal fluctuations implicit in Eq. (9) can disguise quite strong dipole moments as weak susceptibilities: for example, a *p*=1 D electric dipole moment in a practical *E*=100 kV/cm electric field at T=300 K is effectively suppressed by a factor of $pE/3k_BT \approx 3\times 10^{-3}$.

As the vibrational temperature of the cluster decreases below the dipole fluctuation (or anisotropy) energy, the end result is the locked-moment regime, i.e., the problem of a polar rigid body rotating within an external field. The subject is recognizable from the classical and quantum-mechanical treatments of the rotational dynamics of rigid bodies, molecules, nuclei, etc.; the theoretical treatments and results are interestingly varied, and not all questions are yet resolved.

The most important variation arises based on the symmetry of the cluster in question. A rigid top can be characterized by its three principal moments of inertia $I_{a,b,c}$. Often their reciprocals are used: rotational constants *A,B,C* defined, e.g., as $B=\hbar^2/2I_b$, and by convention assigned as *A≥B≥C*.  (The rotational constant *B* should not be confused with the magnetic field denoted by the same symbol.)  If *A=B=C*, the top is called spherical; if *A=B* it is symmetric; and if all constants are different this is an asymmetric top.  As is well known, the free motion of a spherical or symmetric top has an exact solution, while the problem of the asymmetric top is non-integrable and can exhibit unstable dynamics.

For symmetric rotor clusters, in the second order of perturbation theory one finds

$$\langle p_z \rangle = \frac{p^2 E}{k_B T_{rot}} z(\kappa) \tag{10}$$

(and the same for *μ* in place of *p*), where *κ=(C−A)/C* for a prolate top (with *B=C<A* and therefore *κ<0*), and *κ=(A−C)/A* for an oblate top (with *B=A>C* and *0<κ<0.5*), and *z(κ)* is a somewhat bulky function written out in (Bulthuis et al. 2008).  For small deviations from a spherical rotor, i.e., for small *κ*, one finds $z(\kappa) = \frac{2}{9}\left(1 - \frac{1}{5}\kappa + ...\right)$.  For a spherical rotor cluster, *z(0)=2/9* (instead of the universal factor of 1/3 for the Langevin-Debye thermal-equilibrium value).  For deflecting fields stronger than those treatable by perturbation theory, numerical evaluation may become necessary (Dugourd et al. 2001, Ballentine et al. 2007).  From this result one already sees that for extracting reliable values of the dipole moment from a deflection profile, even for a rigid symmetric-top system it is important to have good knowledge of the shape parameter *κ* of the cluster as well as of its rotational temperature.  Otherwise the information obtained will at best qualify as "semi-quantitative."

For asymmetric clusters the situation is not clear-cut.  No analytical solution for adiabatic-entry orientation for asymmetric tops is available, and even a second-order perturbation approach is intractable, hence in general one has to resort to numerical diagonalization of the rotation+field Hamiltonian (Abd El Raheem et al. 2005, Moro et al. 2007), or to classical molecular dynamics simulations which are applicable for large angular momenta (Dugourd et al. 2006).  However, there are indications that out of complexity there may arise simplicity.  Qualitatively (to our knowledge, there are no rigorous proofs of these conjectures), the non-integrability of asymmetric top rotational motion implies a nascent chaotic behavior, which in turn hints at ergodicity and thermodynamics.  As a result, the statistical Langevin–Debye behavior is restored (see Fig. 2 and Sec. 6.2).  This can take place when sufficiently high rotational levels come into play (Bulthuis et al. 2008), in particular when the Stark (or Zeeman) diagram of the rotational levels is replete with avoided crossings (Xu et al. 2008).  The latter has been identified as one of the marks of quantum chaos, and can be enhanced by perturbation of the top by weak collisions with background gas molecules or by rotation-vibrational coupling (Abd El Rahim et al. 2005, Antoine et al. 2006). Full theoretical criteria for the onset of such simplifying behavior have not yet been established, however, so individual deflection patterns of asymmetric nanoclusters need to be assessed on a case-by-cases basis and/or treated numerically.

*4.2. Beam Broadening*

The average shift of the cluster beam discussed above reflects the overall field-induced reorientation of the particles' dipole moments, in other words, the net orientational polarization of the cluster ensemble induced by the deflecting field. Another important observable is the significant amount of (generally asymmetric) broadening displayed by beams of "locked-moment" clusters. Indeed, the very presence of such broadening is taken as a qualitative indicator that the nanoclusters possess a strong "frozen-in" permanent dipole moment (in contrast, as mentioned above, beams of "floppy" clusters tend to deflect uniformly without broadening, appearing as particles with an enhanced polarizability). The broadening originates from the simple fact that particles enter the deflection field region with randomly oriented dipole moments. Those clusters whose dipole has a component along the field will experience a force opposite to those whose moment points against the field, and as a result the beam spreads out and broadens. For weak fields that do not induce a strong reorientation of the cluster population the broadening will be symmetric (equal numbers of moments pointing along and against the field); increasing field intensity is accompanied by a growing polarization of the cluster ensemble and an increasingly asymmetric broadening (as well as an increasing overall shift of the beam's "center of mass" discussed above), see Fig. 3(b).

For weak fields, the broadening of the beam can to first order be taken to be proportional to the zero-field population (i.e., that which exists prior to entering the field region) of various $p_z$ (or $\mu_z$) projections of the cluster dipole moment (see Eq. (8)). For rigid spherically-symmetric tops, for which the rotational energy is given by

$$\varepsilon = BJ(J+1) \qquad (11)$$

($B$ is the rotational constant and $J$ is the angular momentum quantum number) the component of the dipole moment along the $z$ axis is (Townes and Schawlow 1975)

$$p_z = p \frac{MK}{J(J+1)} \qquad (12)$$

($M$ and $K$ are projections of the angular momentum on the space- and body-fixed axes). For a fixed $J$, the population of each $M,K$ sublevel is the same, $P_0(M,K)=1/(2J+1)^2$, and so the probability of $p_z$ having a certain value is proportional to the number of ways that integers $M$ and $K$, each ranging from $-J$ to $+J$, can be combined into a product equal to $J(J+1)(p_z/p)$. Under the assumption that mostly $J \gg 1$ states contribute to the beam population and therefore $J(J+1) \approx J^2$ and $(2J+1)^2 \approx 4J^2$, one finds the following zero-field probability distribution of the dipole moment projections:

$$P_0(p_z) = \frac{1}{2p}\ln\left(\frac{p}{|p_z|}\right) \tag{13}$$

(and similarly for magnetic moments $\mu_z$), see Fig. 3(a). The singularity at $p_z=0$ is just an artifact of the continuum approximation (Bertsch 1994). Since expression (13) is independent of $J$, the entire zero-field moment projection distribution for the whole cluster beam ensemble has the same form.

As stated above, to lowest order the effect of the deflecting field is to displace each dipole according to its original orientation. The weak-field beam profile $b(z)$ is therefore a convolution of the profile of the entering beam $b_0(z)$ and the moment component distribution (13): $b(z) \approx b_0(z) * P_0(p_z)$. Since the variance of a convolution is the sum of two individual variances (Jaynes 2003), and the variance of the distribution (13) is $\sigma_p^2 = p^2/9$ (as can also be derived directly from Eq. (12) (Schäfer, Assadollahzadeh et al. 2008)), it follows that the magnitude of the dipole moment can be deduced from the degree of beam broadening as

$$p = 3\sqrt{\sigma_b^2 - \sigma_{b_0}^2} \,. \tag{14}$$

It is important to keep in mind the relatively narrow region of applicability of this expression: spherically symmetric tops in the weak-field (uniform broadening) regime. For rigid symmetric rotors the low-field $\sigma_p$ becomes a function of the asymmetry parameter $\kappa$ encountered in Eq. (10) (Bulthuis and Kresin, to be published).

For asymmetric tops the response to external fields is dominated by the aforementioned avoided crossing phenomenon, and one finds that beam broadening should decrease with increasing rotational temperature (Xu et al. 2008): $\sigma_\mu^2 \approx \mu^2 B/3k_B T\sqrt{(\gamma-1)}$ (written for magnetic dipoles for which the result was originally derived), where it is assumed that the density of states is proportional to some power of the excitation energy: $\rho(\varepsilon) \sim \varepsilon^\gamma$, which is often the case. Electric-field deflection of asymmetric molecules provides qualitative support to this prediction (Carrera et al. 2008).

*4.3. Résumé*

In summary, then, the deflection pattern of a beam of polar clusters provides two main variables that in principle can be correlated with the value of the dipole moment: the average displacement and (for locked-moment species) the amount of broadening. For certain families and in certain limits, these correlations can be derived analytically; for other situations, such as asymmetric-top particles and strong deflecting fields, one has to resort to numerical simulation of the beam profile. The analytical expressions play another important role: they point out the fact that unless one begins with a consistent physical picture of the cluster's shape, structure, and temperature (rotational or vibrational, as required by the situation), quantitative errors can result from extracting its electric and magnetic dipole moments from field deflection profiles.

## 5. Electric Dipole Moments in Selected Cluster Families

This section outlines experimental information on the electric dipole moments of several selected cluster families obtained via beam measurements. (An analogous overview of magnetic moments of free clusters is presented in the next section.)

*5.1 Metal-Fullerene Clusters*

A beautiful symmetric-top polar system illustrating "textbook" response behavior ranging from the floppy to the rigid is a fullerene $C_{60}$ with a metal atom (usually an alkali) riding on the outside of the cage (Fig. 4). These beam-deflection experiments are reviewed in detail by Broyer et al. (2002, 2007). Because of charge transfer from the atom to the $C_{60}$ cage, a strong dipole is formed.

At low temperatures the alkali atom (and hence the dipole moment) is bound to the cage and the electric deviator induces significant beam broadening as shown in Fig. 5(a). Analogous behavior is found for the strongly bound $TiC_{60}$ (Dugourd et al. 2001). As the temperature increases, the alkali atom becomes able to move freely around the cage (the hopping barrier has been found to be 0.02 eV (Dugourd et al. 2000)), and the beam profile changes into a pure deflection pattern, Fig. 5(b). The dipole moments of $MC_{60}$, determined from the beam profiles, increases with the size (and hence decreasing ionization potential) of the alkali atom M, ranging from 12.4 D for M=Li to a huge 21.5 for M=Cs.

The experiments were extended to fullerenes carrying not just one, but a whole cluster of alkali molecules on the surface ($Na_nC_{60}$).

Interestingly, it appears that for *n* up to eight the atoms are spread around on the surface of the cage ("wetting"), but larger sodium clusters tend to collect into a droplet ("non-wetting"), see Fig. 6. For understanding the dipole moments of the wetted picture, it is important to keep in mind that whereas in the lowest-energy state the Na atoms may arrange themselves symmetrically around the cage and thereby produce only a low net dipole moment, at finite temperatures they slide around on the surface which gives rise to a much higher dipole. Thus the time-averaged dipole moment that enters Eq. (9) is itself temperature-dependent (Broyer et al. 2007), an effect that may be labeled "vibronic-induced polarization."

*5.2 Water Clusters*

The high dipole moment of the isolated water molecule, 1.855 D (Clough et al. 1973), which in a liquid is further enhanced to approximately 3 D (at *T*=300 K) by intermolecular interactions (Gubskaya and Kusalik 2002), is responsible for the chemical and solvent properties of water and, as one corollary, for the form of life as we know it. Despite its relatively simple molecular structure, water in the condensed phase presents a major challenge to theory and simulations because of its very high rotational and hydrogen-exchange mobility and the propensity to form a network of crucially important but fleeting and fluctuating hydrogen bonds. Water clusters have been extensively studied as model systems for understanding such correlated networks, and the dipole moment of a nanodroplet or nanoicicle of a finite number of water molecules is an important characteristic of such an assembly.

High-resolution laser measurements of the Stark effect of so-called vibration-rotational tunneling spectra of very cold clusters in a beam were performed by Gregory et al. (1997). They obtained an electrical dipole moment of ≈1.9 D for the cage isomer of the $(H_2O)_6$ species, in agreement with calculations (see Fig. 7).

Beam deflection experiments were carried out by Moro et al. (2006). In this case the clusters were estimated to be at a temperature of approximately 200 K, and the beam exhibited a uniform deflection proportional to the square of the electric field, a picture fully consistent with orientational polarization of a "floppy" dipole. Using Eq. (9) and subtracting the electronic polarizability contribution, values of *p*≈1.3 D for $(H_2O)_{3-8}$ and p≈1.6 D for $(H_2O)_{9-18}$ were obtained. In this situation, the *p* value represents the time-averaged total dipole moment of a nano-assembly of rotationally labile polar water molecules. Repeating the measurement under colder expansion conditions yielded $\chi$ values shifted upwards in agreement with Eq. (9), implying that the average magnitude of *p*, and therefore the landscape of molecular fluctuations, remained unchanged down to ~120 K,

12and therefore suggesting that at this temperature there still had been no freezing transition to a state of molecular orientational ordering .

### 5.3 Alkali Halide Clusters

The alkali-halide salts are prototypes of ionically bonded polar structures. In fact, the very first electrostatic beam deflection experiment was performed in 1927 on the KI molecule (Wrede 1927, Ramsey 1956). The primary structural motif of alkali-halide clusters is believed to be that of cubic nanocrystals analogous to the bulk lattice structure. Deflection experiments have been performed on a number of $M_nX_{n-1}$ clusters ($M$=alkali atom, $X$=halogen atom) with $n$ in the range from 2 to 20 (Rayane et al. 2002, 2003; Jraij et al. 2006). Many of these are considered to be cuboids with a vacancy, in particular a corner vacancy, which is occupied by the electron contributed by the excess metal atom (see Fig. 8). This gives rise to strong dipole moments, $p$~3-10 Debye.

An interesting feature of the ionic nanocrystal data is that in the temperature range from 77 K up to 500 K, the cluster deflection profiles are uniformly displaced, unbroadened, and with an average shift proportional to the square of the electric field, corresponding to a susceptibility of the form given in Eq. (9). This looks counterintuitive, because these clusters would be expected to be of the "rigid dipole" type. The response dynamics of these systems is not fully understood, although simulations suggest that it is strongly affected by the presence of vibrational excitations (Rayane et al. 2002), and it would also be interesting to explore the implications of the avoided-crossing model (Xu et al. 2008).

### 5.4 Metal Cluster Dipole Moments and Metal Cluster Ferroelectricty

One of the most characteristic properties of metals is that they cannot support voltage differences. This implies the absence of electric fields in bulk metals. Noting that even a relatively small dipole moment of 1 D will produce local electric fields of the order of $10^7$ V/cm in a 1 nm particle, one might expect that electric dipole moments are rare in metal clusters. This is indeed found to be the case. With very few exceptions, the electric dipole moments in metal clusters are immeasurably small («1D). For example, recent measurements of sodium clusters up to $Na_{200}$ show that only $Na_3$ has a weak dipole moment of ≈0.02 D, for all other clusters the dipole moments are even smaller (Bowlan et al., to be published). Apparently even very small clusters behave like bulk metal particles which cannot have electric dipole moments.

The general absence of dipole moments in metal clusters draws attention to those few species that violate this rule. For example, rhodium clusters in a beam, $Rh_{5-28}$, have been investigated by Beyer and Knickelbein (2007). Most exhibited small uniform deflections in the high field direction, corresponding to purely polarizable behavior. However, $Rh_7$ and $Rh_{10}$ manifested anomalously high susceptibility values, suggesting the presence of permanent dipole moments. The $Rh_7$ beam behaved as that of a rigid rotor, displaying relatively symmetric broadening and a centroid shift compatible with Eq. (10). Using this equation with the assumption $\kappa \approx 0$ yielded an electric dipole moment of 0.24 D. $Rh_{10}$, on the other hand, differed in producing only a uniform shift towards the high field, and in being sensitive to variation of cluster vibrational temperature by the source expansion conditions. This behavior was interpreted as being characteristic of a spatially fluctuating (i.e., floppy) dipole. The suggestion of two closely related nanoparticles displaying two different types of dipole moment coupling is interesting and highlights the sensitivity of cluster polarization to cluster structure.

A somewhat analogous picture emerged in a study of lead clusters $Pb_{7-38}$ by Schäfer, Heiles et al. (2008). In this case, $Pb_{12,14,18}$ exhibited anomalous deflections which were interpreted as evidence for their possessing a polar structure (and in the case of $Pb_{12}$ as there being two isomer populations in the beam, one polar and the other non-polar).

Besides these few and rare cases, there is also an entire class of metal clusters that display unusual polarization. Clusters of vanadium, niobium and tantalum, as shown by Moro et al. (2003), as well as several alloys of these metals, behave normally at room temperature, but exhibit very large dipole moments at cryogenic temperatures (typically below 25 K). The origin of this effect if very different from the electric dipole moments of non-metallic clusters. The fact that they only have dipole moments at low temperatures and not at higher temperatures, indicates that they undergo a phase transition from a normal state to one that has an electric dipole moment. Materials with this property are called ferroelectrics.

In the bulk, ferroelectric materials are always non-metals. Hence, this form of metallic ferroelectricity is a cluster property. Moreover, not only are the dipole moments large, they are ubiquitous and in all cases show several distinctive properties (Xu et al. 2007). One is that the dipole moment has an odd-even effect. That is, the dipole moments of the even clusters are systematically larger than those of the adjacent odd clusters. Neither the dipole moment nor the even-odd effect show a tendency to diminish with increasing cluster size. Moreover, careful experiments in which "impurity" atoms were imbedded in the clusters, demonstrated that the odd-even effect is directly related to the number of electrons in the cluster. These unusual effects indicate that the electric dipole moments are not directly related to the atomic structure of the clusters, but rather to the electronic structure of the materials. Further measurements showed that the dipole moments themselves are not rigidly attached to the cluster framework, which further supports this interpretation.

A full understanding of the nature of ferroelectricity in metallic clusters is still lacking, however it is intriguing to take note of a correlation between metals that exhibit the ferroelectric effect and those that are superconductors in the bulk.

*5.5 Extremely Polar Chains Assembled in Helium Nanodroplets*

Finally, we ought to mention a different kind of nanocluster assembly, namely one that involves polar impurities in helium nanodroplets. The technique involves atoms and molecules picked up, entrapped, transported, and cooled to sub-Kelvin temperatures by a beam of superfluid $^4\text{He}_n$ nanodroplets ($10^2 \lesssim n \lesssim 10^5$) generated by supersonic expansion of pure helium gas through a small cryogenically cold nozzle (Toennies and Vilesov 2004, Stienkemeier and Lehmann 2006).

A unique feature of this approach is that the picked-up molecules are very rapidly quenched to the nanodroplet temperature, which is approximately 0.37 K. As a result, very interesting metastable arrangements with unusual properties can be produced and "frozen" into long-term existence. Nauta and Miller (1999) discovered that upon sequential pick-up, linear polar HCN molecules ($p_{HCN}$=3.0 D) are steered by long-range electrostatic (dipole-dipole) forces to self-assemble into long chains, aligned head-to-tail inside the nanodroplet. An analogous observation was made for HCCCN, $p_{HCCCN}$=3.7 D (Nauta et al. 1999) The length of the chain seems to be limited only by the droplet diameter. In normal environments such a system would be unstable, as the orientation effects would be weak relative to the large thermal rotational energy of the molecules. But within the very cold and inert liquid helium droplet this becomes a long-lived configuration.

The structure of the chains was verified by high-resolution vibrational spectroscopy (Fig. 9) of doped droplets passing through a region of strong uniform electric field. The field enforces so-called "brute force" orientation of the polar impurities, so that their rotations are replaced by pendulum-like oscillations about the field direction ("pendular state spectroscopy").

Very interesting aspects of these systems are the facts that these are electric dipoles of unprecedented magnitude in the molecular and cluster research fields (e.g., a chain of seven lined-up HCN molecules will have a dipole moment of as much as ≈20 D, and possibly even higher due to enhancement by nearest-neighbor interactions), and that by virtue of being embedded in the superfluid nanodroplet, the chains are not only aligned, but also vibrationally and rotationally cold.

# 6. Cluster Ferromagnetism

## 6.1 Ferromagnetism

Ferromagnetism is the bulk property in which a material spontaneously acquires a magnetic moment (for an introduction see, e.g., (Kittel 1953)). Ferromagnets are typically metals and the magnetic moment is associated with a spontaneous alignment of the electronic spins. This is mediated by an interaction between the electrons which favors mutual alignment of the spins. Physically, this tendency reflects a closely related atomic property where electrons in close proximity tend to align their spins in order to minimize the wave function overlap and thereby to minimize their mutual electrostatic repulsion. Hence, spin alignment tends to reduce the potential energy of the system. This somewhat subtle property of electrons is called the exchange interaction.

The tendency to reduce the potential energy comes at the cost of increasing the kinetic energy of the electrons. In a dilute electron gas, electrons favor a spin up-spin down arrangement resulting in no net magnetic moment. The reason is that this configuration minimizes the total kinetic energy: electrons can occupy the same electronic orbits if their spins are opposite. In contrast to a dense electron gas, in a dilute electron gas the resulting large overlap of the electronic clouds does not significantly increase the potential energy.

In all metals itinerant valence electrons spend some time near the atomic cores where the electronic densities are relatively high and they interact with each other rather strongly, so that the spins tend to align. They also spend time between the cores where the electronic density is low and here the low spin configuration is favored. In ferromagnets, the tendency to align the spins dominates so that the metal spontaneously polarizes.

The magnetic moment that arises from ferromagnetism has no unique preferred direction. In some cases the magnetic moment direction is isotropic so that it can point in any direction. In other cases it has a weak tendency towards a preferred axis, in this situation the magnetic moment may be either aligned or anti-aligned with that direction. Consequently, in contrast to those electric dipole moments that are rigidly attached to the geometrical structure, the magnetic dipole moment is relatively free to change its direction while keeping its magnitude constant.

At sufficiently high temperatures (the Curie temperature of the material) the magnetic moment vanishes. At this point thermal vibrations become competitive with the energies associated with the mutual alignment of the spins, thereby reducing and ultimately extinguishing the magnetic moment.

*6.2 Single-Sided Deflections of Ferromagnetic Clusters*

Ferromagnetism also manifests itself in clusters of ferromagnetic metals, as first demonstrated by de Heer et al. (1990). The most striking property of free ferromagnetic clusters is that their magnetic moments appear to spontaneously align with an applied magnetic field. The effect reveals itself as single-sided deflection of a beam of ferromagnetic clusters in an inhomogeneous magnetic field (a Stern-Gerlach magnet), in contrast to the expected symmetric deflections of paramagnetic clusters (as seen, for example, in the classical Stern-Gerlach experiment on silver atoms). This property is shown in Fig. 10, contrasting Stern-Gerlach deflections of paramagnetic atoms with those of paramagnetic clusters.

For clusters it is experimentally found (see Fig. 2) that the average magnetization $\langle M_z \rangle$ of the clusters in weak magnetic fields corresponds to

$$\langle M_z \rangle = \frac{\mu^2 B}{3 k_B T}, \qquad (15)$$

where $\mu$ is the magnetic moment, $T$ is the temperature of the cluster source, and $B$ is the magnetic field strength, reminiscent of the low-field limit of the Langevin equation. These deflections are observed for all clusters, even at cryogenic temperatures. However, as emphasized in Sec. 4.1, the clusters are not in contact with a heat bath since they are isolated when they pass through the magnet. Moreover, many are small and cold enough that an internal statistical temperature cannot be defined. For example, a cluster with 5 atoms emerging from a cluster source with $T=20$ K contains on average much less than one phonon (the quantum of lattice vibrations).

Thus this limit corresponds to the aforementioned case of asymmetric rigid dipolar species. As discussed by Xu *et al.* (2008), the single-sided deflections ultimately result from alignment of the magnetic moment of each specific cluster in a magnetic field. When a cluster with a magnetic moment is placed in a magnetic field, it will precess much like a spinning top, and this will cause the magnetic moment on average to slightly align with the magnetic field. The degree of alignment depends on the rotational state of the cluster. Each individual cluster rotates at a different rate that follows a distribution given by the temperature in the source. In the case of a dense spectrum, it has been shown by Xu et al. (2008) that quite generally the average deflection of the entire cluster ensemble will be given by Eq. (15) for a wide range of cluster sizes and temperatures.

*6.3 Ferromagnetic Cluster Magnetic Moments*

Cluster magnetic moments can be extracted from the beam deflection data using Eq. (15). They have been found by Billas et al. (1994) to be significantly larger than the expected moments extrapolated from the bulk (Fig. 11). The enhanced magnetic moments were explained in terms of an enhancement caused by the relatively large surface area in clusters, since it is known that the magnetic moments at ferromagnetic surfaces are enhanced. Specifically, for small clusters of iron, cobalt and nickel the magnetic moments correspond approximately to 3, 2, and 1 $\mu_B$ per atom. These values are readily understood since they correspond to the total spin of the 3d electronic shells of the corresponding atoms. As the clusters increase in size, the magnetic moments decrease towards the corresponding bulk values.

Using the molecular beam deflection technique, the magnetic moments of a variety of ferromagnetic clusters have been measured, including alloys (see, e.g., (Yin et al. 2007)). Particularly interesting are metals which are non-magnetic in the bulk but of which the clusters are ferromagnetic. Rhodium clusters (Cox et al. 1993) are an important example of this class of ferromagnetic clusters that also exhibit unusual electric dipole polarizabilities as mentioned above.

## 7. Summary

Magnetic and electric dipole moments of cluster nanoparticles are observables displaying high sensitivity to the electronic and geometric structure of the system. Therefore they are of considerable experimental and theoretical value in nanoscience. It is very appealing that both the intrinsic magnitudes of these quantities and their experimental manifestations (e.g., in beam deflection studies and in spectroscopy) involve a wide range of fundamental issues, for example phase transitions, exchange interactions, long-range forces, chemical bonding, quantum and classical chaos, superconductivity and superfluidity, etc. At the same time, one has to be cognizant of the fact that quantitative interpretation of the experimentally measured patterns requires careful awareness of the underlying assumptions and mechanisms.

## Acknowledgements

Useful discussions with J.Bowlan, J.Bulthuis and A.Liang, and support from the U.S. National Science Foundation are sincerely appreciated.

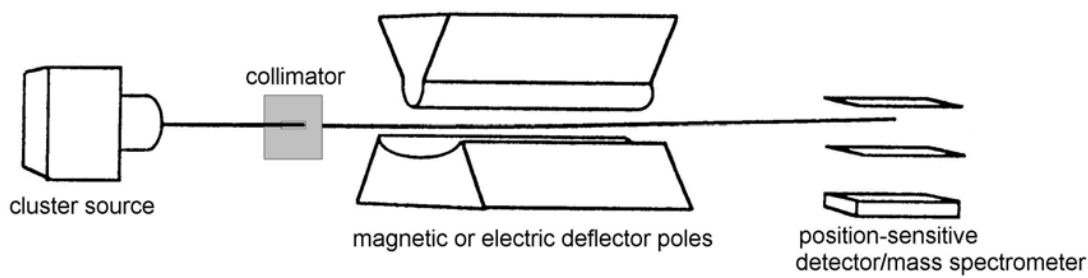

**Fig. 1.** Outline of a configuration for electric or magnetic deflection measurements for nanocluster beams.

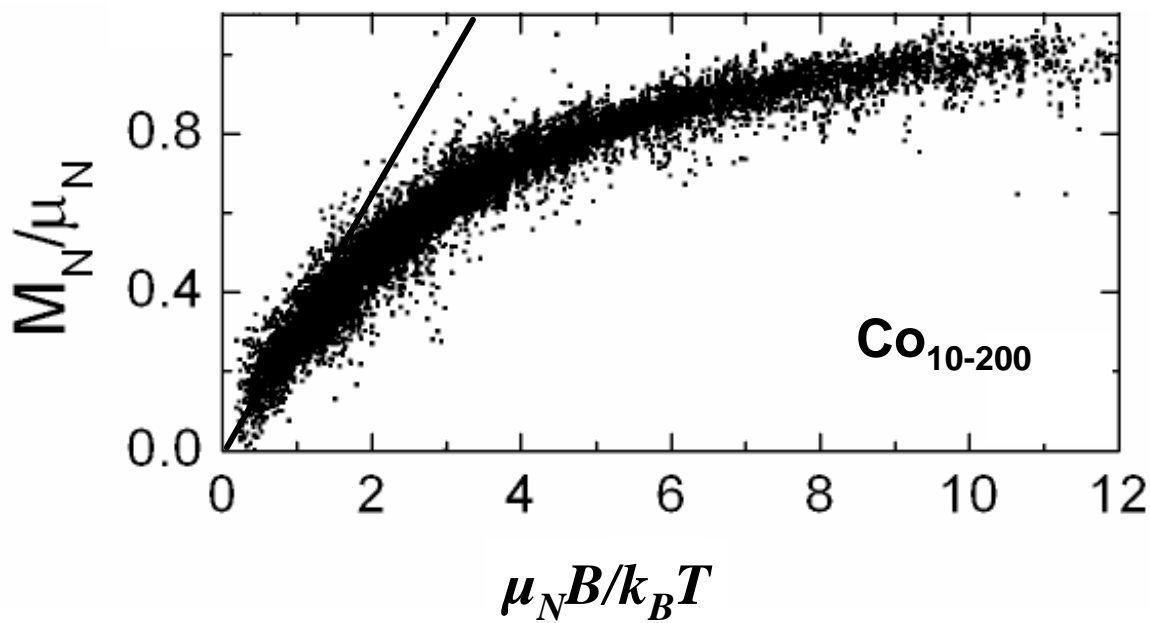

**Fig. 2.** Normalized magnetization projections of Co$_N$ clusters ranging in size from $N=10$ to 200, plotted as a function of $x = \mu_N B / k_B T$, from Xu et al. (2008). The plot combines 63 separate data sets collected at nine temperatures from 25 to 100 K and seven fields from 0 to 2 T, representing about 10 000 data points in total, all collapsing onto a universal curve. The trend is consistent with the Langevin function, although the latter approaches saturation slightly slower than the data points. The initial slope is consistent with Eq. (15). The underlying mechanism in this case is not internal spin relaxation but rotational-precessional dynamics, and the relevant $T$ is the rotational temperature of the cluster beam.

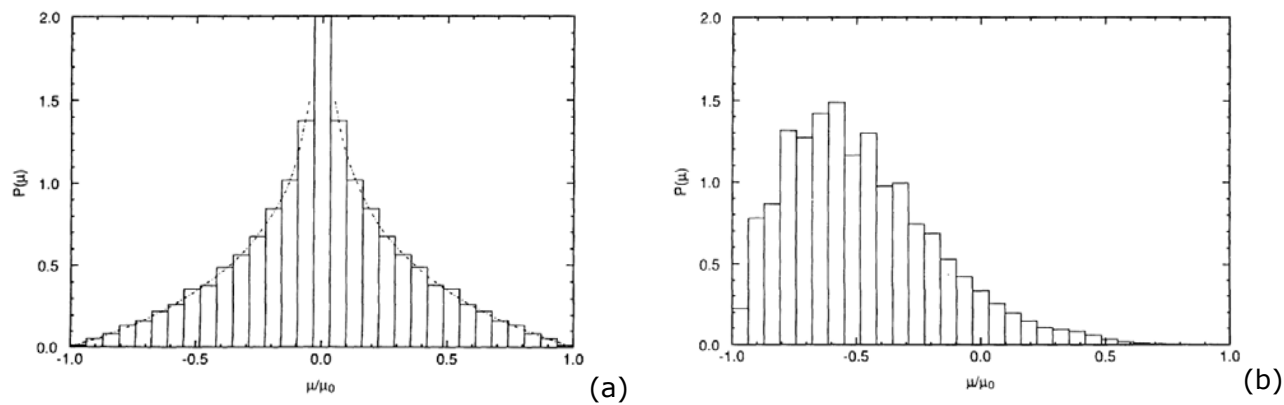

**Fig. 3.** Distributions of magnetic moment projections for an ensemble of spherically symmetric rigid rotors with a magnetic moment $\mu_0$, from Bertsch and Yabana (1994). The histograms show numerical results for states with $J_{max} \leq 40$ and rotational temperature $T_{rot}=100B$, where $B$ is the cluster rotational constant. (a) Zero magnetic field; the dashed line is the approximation Eq. (13). (b) Adiabatic entry into a magnetic field of intensity equal to $4T_{rot}/\mu_0$.

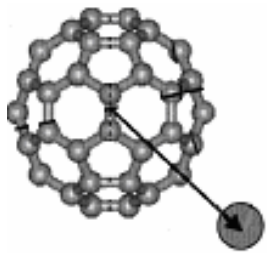

**Fig. 4.** A polar metal-fullerene cluster (Rayane et al., 2003).

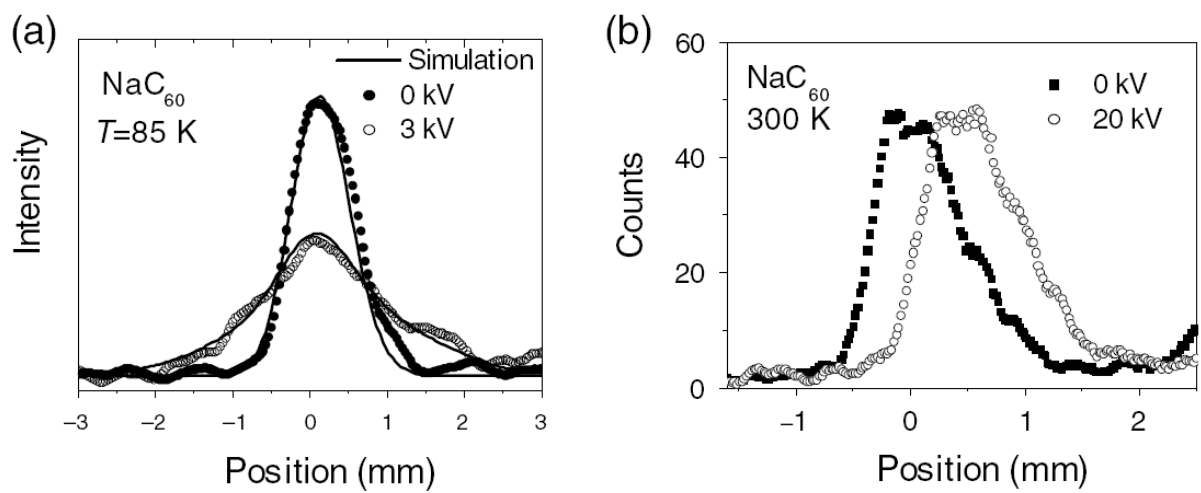

**Fig. 5.** Experimental (circles) beam profiles for the deflection of NaC$_{60}$, (a) at 85 K and (b) at 300 K, from Broyer et al. (2007). On the left, the solid line is a calculation for a rigid locked-moment rotor with $p=14.8$ D. On the right, the deflection pattern reflects a fluctuating dipole resulting from the mobility of the alkali atom on the fullerene cage.

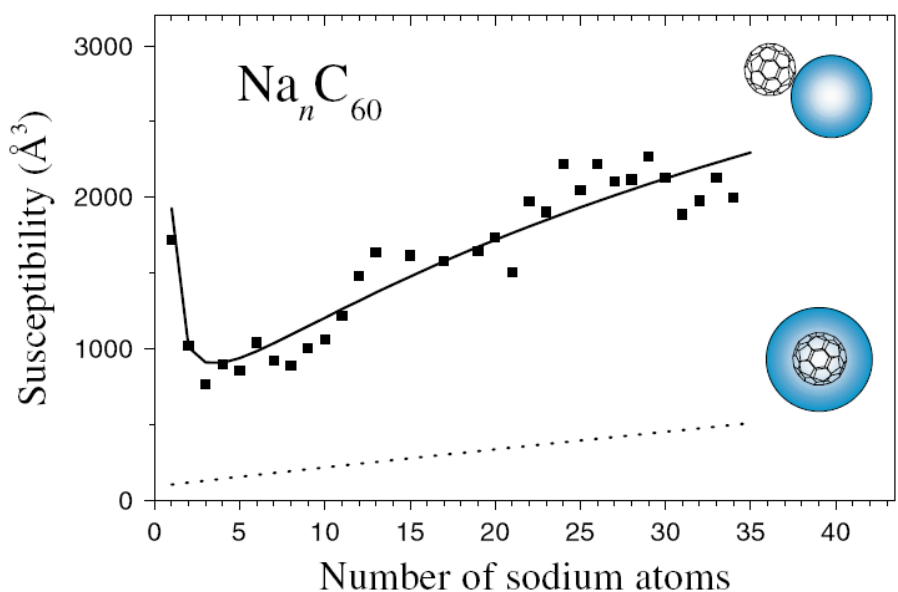

**Fig. 6.** Susceptibility of Na$_n$C$_{60}$ clusters, from Broyer et al. (2007). Dark squares: experimental values. Dashed line: values calculated assuming that the sodium atoms form a metal shell around the fullerene. Full line: values calculated assuming formation of a metal droplet on the surface.

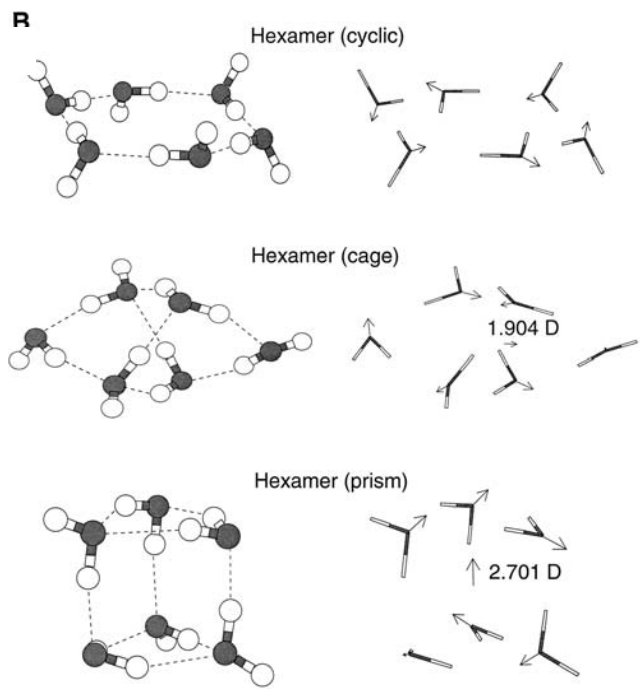

**Fig. 7.** Three theoretical structures of the water hexamer and their dipole moments (Gregory et al. 1997). (The cyclic isomer has no dipole moment.)

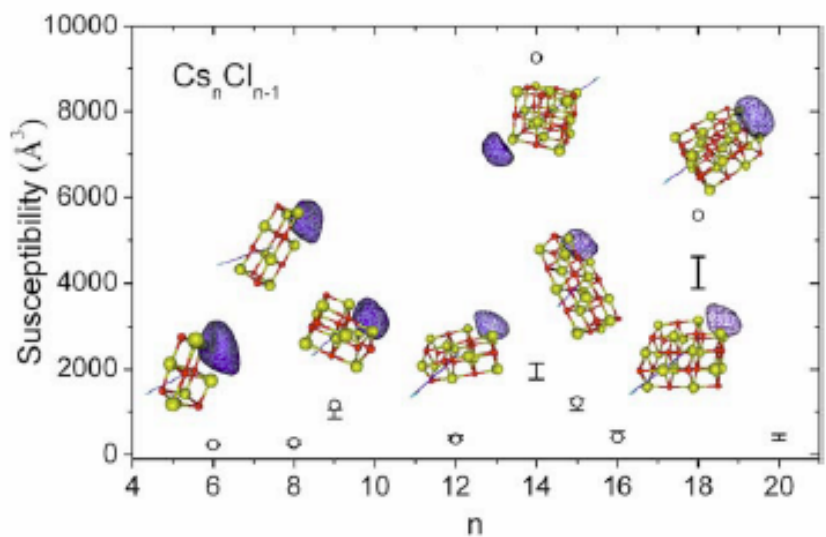

**Fig. 8.** Calculated (circles) and experimental (bars) electrical susceptibilities of $Cs_nCl_{n-1}$ clusters, from Jraij et al. (2006). For each size the theoretical lowest-energy structure is also shown, together with the direction of the electric dipole and a surface of constant charge density for the excess electron.

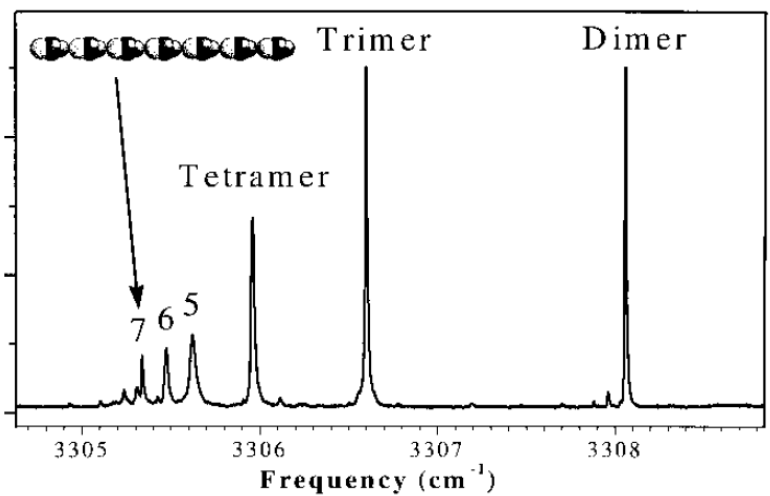

**Fig**. **9**. Spectrum of the free C-H stretching region of HCN chains in helium nanodroplets, showing peaks corresponding to clusters up to at least the heptamer (*n=7*), from Nauta and Miller (1999). A highly polar linear heptamer chain is shown in the inset.

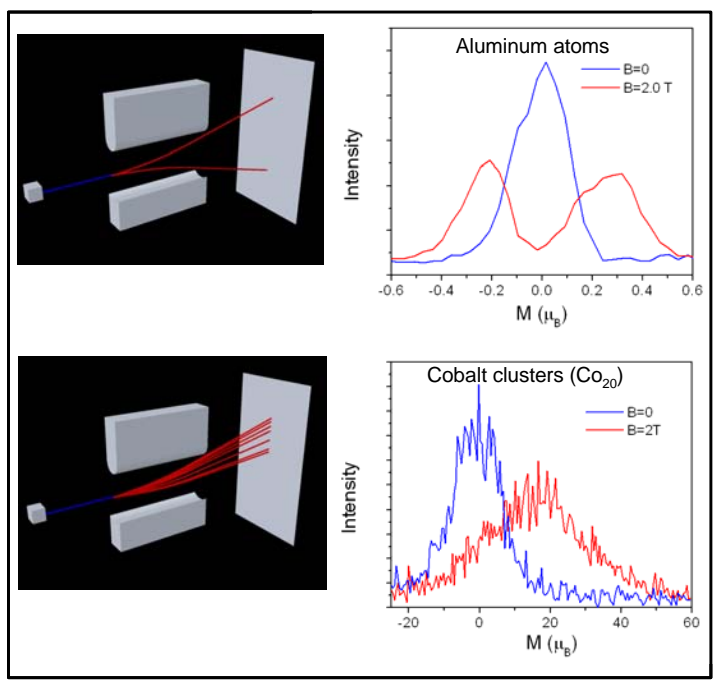

**Fig**. **10**. Stern-Gerlach deflections of atoms and clusters. Aluminum atom deflections exhibit two peaks corresponding to the possible spin orientation with respect to the magnetic field, as shown schematically at top left and experimentally at top right. Ferromagnetic clusters, on the other hand, exhibit single-sided deflections (bottom panels).

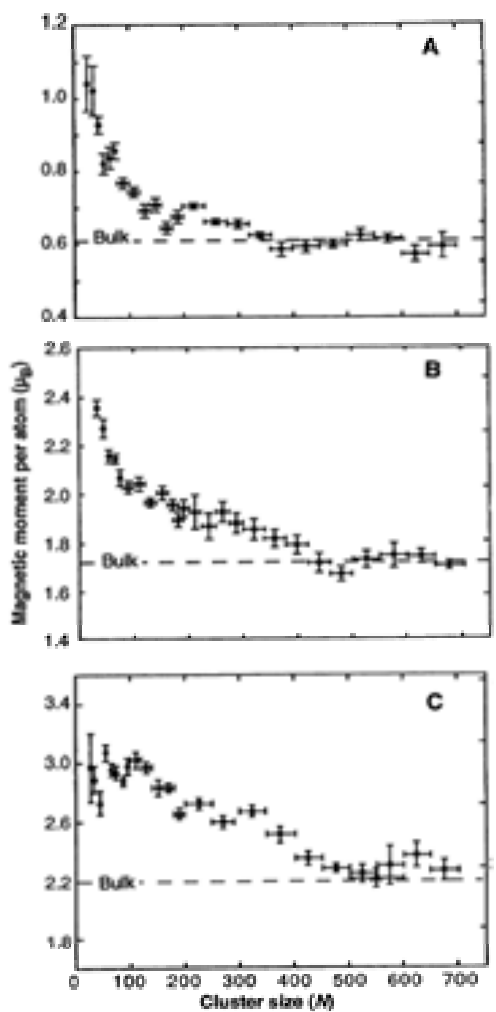

**Fig. 11**. Magnetic moments of (A) nickel, (B) cobalt, and (C) iron clusters as a function of cluster size. For small clusters, the magnetic moments per atom are approximately 1, 2 and 3 $\mu_B$., respectively.